\documentclass[a4paper]{aa}
\usepackage{graphicx}
\def\integral{{\it INTEGRAL}}
\def\chandra{{\it Chandra}}

\def\xmm{{\it XMM-Newton}}
\def \src {XMMU~J174716.1--281048}
\def \igr {IGR~J17464--2811}

\def \hcm {\hbox {\ifmmode $ atom cm$^{-2}\else atom cm$^{-2}$\fi}}

\def\approxgt{\mathrel{\hbox{\rlap{\lower.55ex \hbox {$\sim$}}
        \kern-.3em \raise.4ex \hbox{$>$}}}}
\def\approxlt{\mathrel{\hbox{\rlap{\lower.55ex \hbox {$\sim$}}
        \kern-.3em \raise.4ex \hbox{$<$}}}}

\begin{document}
\title{\src: a ``quasi-persistent'' very faint X--ray transient?}


   \author{M. Del Santo\inst{1} 
        \and L. Sidoli\inst{2}
        \and S. Mereghetti\inst{2}
        \and A. Bazzano\inst{1}
        \and A. Tarana\inst{1,3}
        \and P. Ubertini\inst{1}
        }

\offprints{Melania Del Santo (melania.delsanto@iasf-roma.inaf.it)}
\institute{ Istituto di Astrofisica Spaziale e Fisica Cosmica di Roma -- INAF, via del Fosso del Cavaliere 100, 40133 Roma, Italy
\and Istituto di Astrofisica Spaziale e Fisica Cosmica di Milano -- INAF, via E. Bassini 15, 20133 Milano, Italy
\and Universit\'a di Tor Vergata, Roma, Italy
}

\date{Received .... 2007; Accepted:    }

\authorrunning{M. Del Santo et al.}

\titlerunning{{The very faint X--ray transient \src}}

\abstract
{The X--ray transient \src\ was serendipitously discovered with \xmm\ in 2003. 
It lies about 0.9 degrees off the Galactic Centre and its spectrum shows
a high absorption ($\sim$8$\times$10$^{22}$cm$^{-2}$).
Previous X--ray observations of the source field performed in 2000 and 2001 
did not detect the source, indicative of a quiescent emission at least two orders of magnitude fainter. 
The low luminosity during the outburst ($\sim$5$\times$10$^{34}$ erg s$^{-1}$ at 8 kpc)
indicates that the source is a member of the ``very faint X--ray transients'' class.
On 2005 March 22$^{nd}$ the \integral\ satellite caught a possible type-I X--ray burst from 
the new \integral\ source \igr, classified as fast X-ray transient.
This source was soon found to be positionally coincident, within the uncertainties, with \src.
Here we report data analysis of the X--ray burst observed 
with the IBIS and JEM-X telescopes and confirm the type-I burst nature. 
We also re-analysed \xmm\ and \chandra\  archival observations of the source field.
We discuss the implications of these new findings, particularly related to the source distance
as well as the source classification.
}
\maketitle
%

\section{Introduction}
\label{sect:intro}

\src\ is a faint X--ray transient serendipitously 
discovered in 2003 with \xmm\ in the Galactic Centre (GC) region,
during a pointed observation on the composite SNR G0.9+0.1 (Sidoli \& Mereghetti 2003; Sidoli et al. 2004). 
The observed flux was 3.7$\times$10$^{-12}$~erg~cm$^{-2}$~s$^{-1}$
(2--10 keV), and the spectrum was fit well with an absorbed power-law
model with photon index $\sim$2 and N$_{H}$$\sim$9$\times$10$^{22}$~cm$^{-2}$.
The high interstellar absorption suggested a source location at the GC. 
The derived luminosity, assuming  d=8 kpc, is 5$\times$10$^{34}$~erg~s$^{-1}$.
The source position was also within the  EPIC field of view 
during the \xmm\ pointed observation of SAX~J1748.2--2808 (\cite{sidoli:06}) performed in 2005.
\src\ was imaged at a large off-axis angle ($\sim$13$'$) with an
observed 2--10~keV flux of $\sim$2$\times$10$^{-12}$~erg~cm$^{-2}$~s$^{-1}$ and
a spectrum similar to that observed in 2003.

The discovery of a possible type-I X--ray burst
at the coordinates RA(J2000)=266.810$^{\circ}$, Dec(J2000)=--28.185$^{\circ}$ 
(with a 90\% error radius of 1 arcmin), 
has been recently reported by Brandt et al. (2006) with the 
JEM-X monitor (3--30 keV) on board \integral\ satellite.
The new burster was initially designated as IGR J17464--2811.
Taking into account both the spatial coincidence and
the temporal closeness of the \integral\ (March 2005) and \xmm\ observations
(outburst observed on 2005, 26-27 February),
Wijnands (2006) suggested that the X--ray burst is indeed associated with the transient \src.
 
Here we report results of the \integral\ observation of the type-I X--ray burst.
We also discuss all archival X--ray observations 
of the source field, performed with \xmm\ and \chandra\ satellites.

\section{Observation and data analysis}
\label{sect:obs}

We present \integral\ public data collected with the two coded mask telescopes
JEM-X (\cite{lund:03}) and IBIS (\cite{ube:03}).
In particular, we analysed data of the low energy detector layer of IBIS, ISGRI (\cite{lebrun:03}),
and JEM-X1 camera with OSA 5.1.
The X-ray burst occurred on March 22$^{nd}$ at 07:55:33 UT.
Source light curves during the corresponding \integral\ pointing (lasting 1800 s)
have been extracted in three energy ranges, 3-6, 6-10, 18-26 keV, with 3-second bin-size.
In order to extract both JEM-X and IBIS/ISGRI burst spectra, as well as IBIS images, 
we selected the time interval t$_{start}$=07:55:33 and t$_{stop}$=07:56:52.

\xmm\ (\cite{ja:01}) observed the source
field three times: on 2000, September 23 (as part of the GC monitoring),
on 2003, March 12 (pointed on the SNR G0.9+0.1), and on 2005, February 26-27.
EPIC data have been reprocessed with the version 6.5 of the Science Analysis
Software (SAS) and known hot (or flickering) pixels and electronic
noise, as well as proton flares, were rejected (details on the data reduction and analysis
in Sidoli et al. 2004 and Sidoli et al. 2006).

The source field has also been observed with $Chandra$/ACIS
on 2000, October 27 (pointed on the SNR G0.9+0.1)
and on 2001, July 16 (Obs.ID~2271 and 2274; Wang et al. 2006). 
The events files (level 2) processed by the \chandra\ X--ray Centre and available from
the public archive have been analysed by using CIAO tool v. 3.2.2.

   \begin{figure}[t!]
   \centering
   \includegraphics[width=4.5cm]{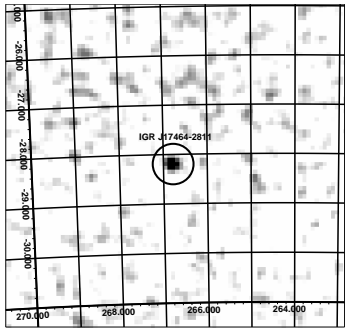}
      \caption{\src/IGR J17464-2811 observed with IBIS/ISGRI between 20 and 25 keV. 
The image has been collected in 63 seconds
        during the X-ray burst.}
         \label{fig:ibis_ima}
   \end{figure}
%

   \begin{figure}[t!]
   \centering
     \includegraphics[width=5.75cm]{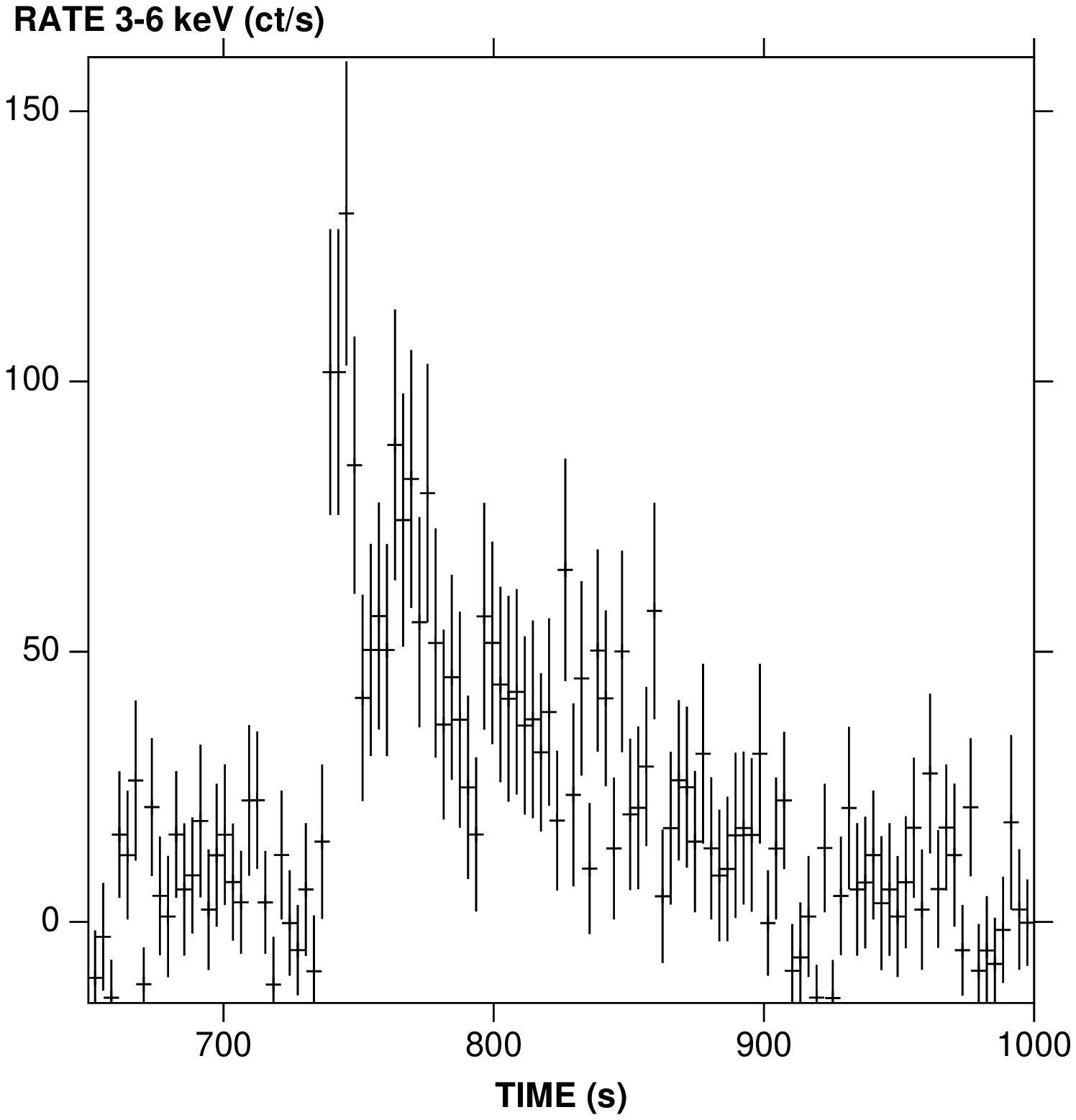}
    \includegraphics[width=5.65cm]{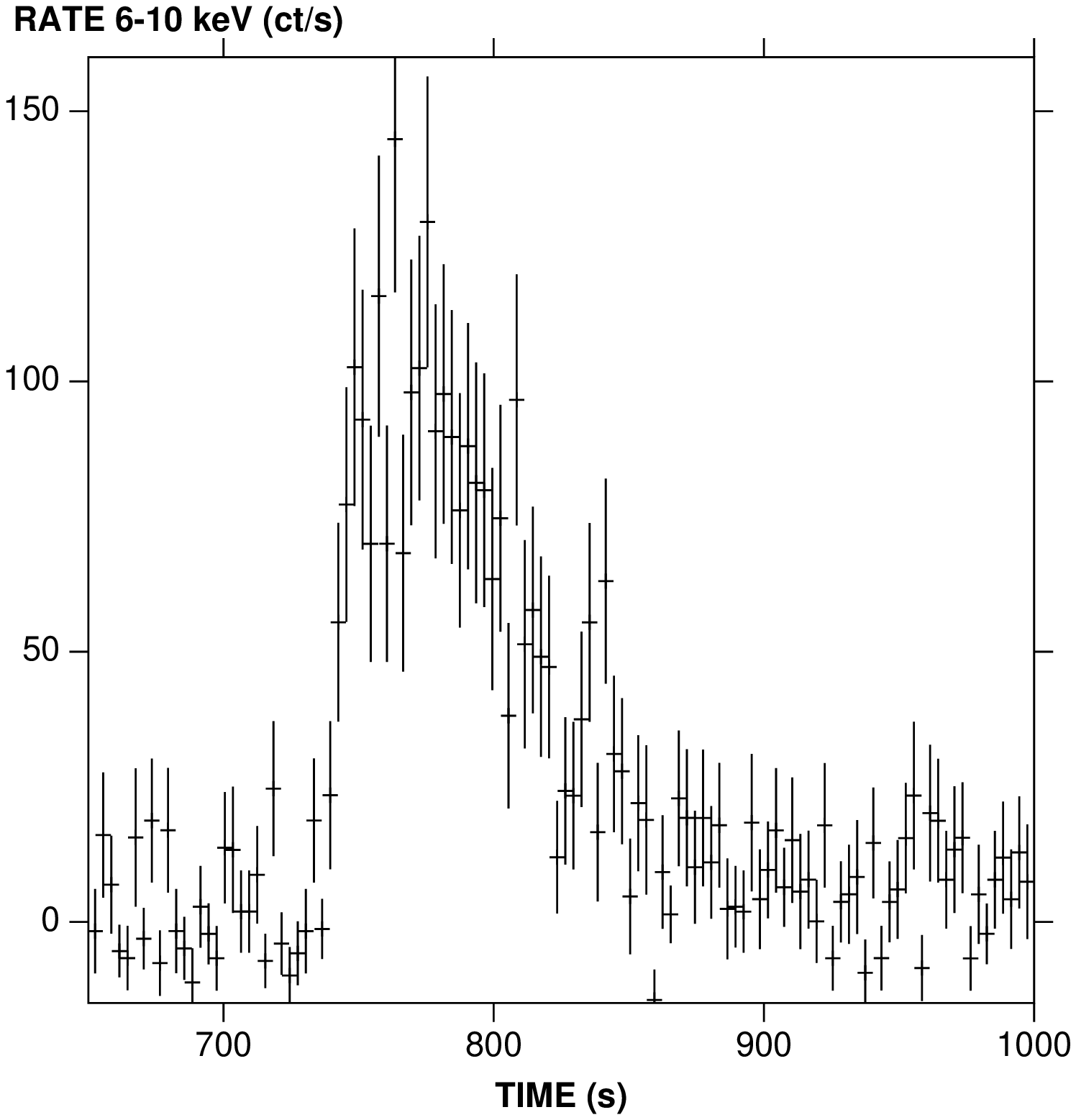}
     \includegraphics[width=5.65cm]{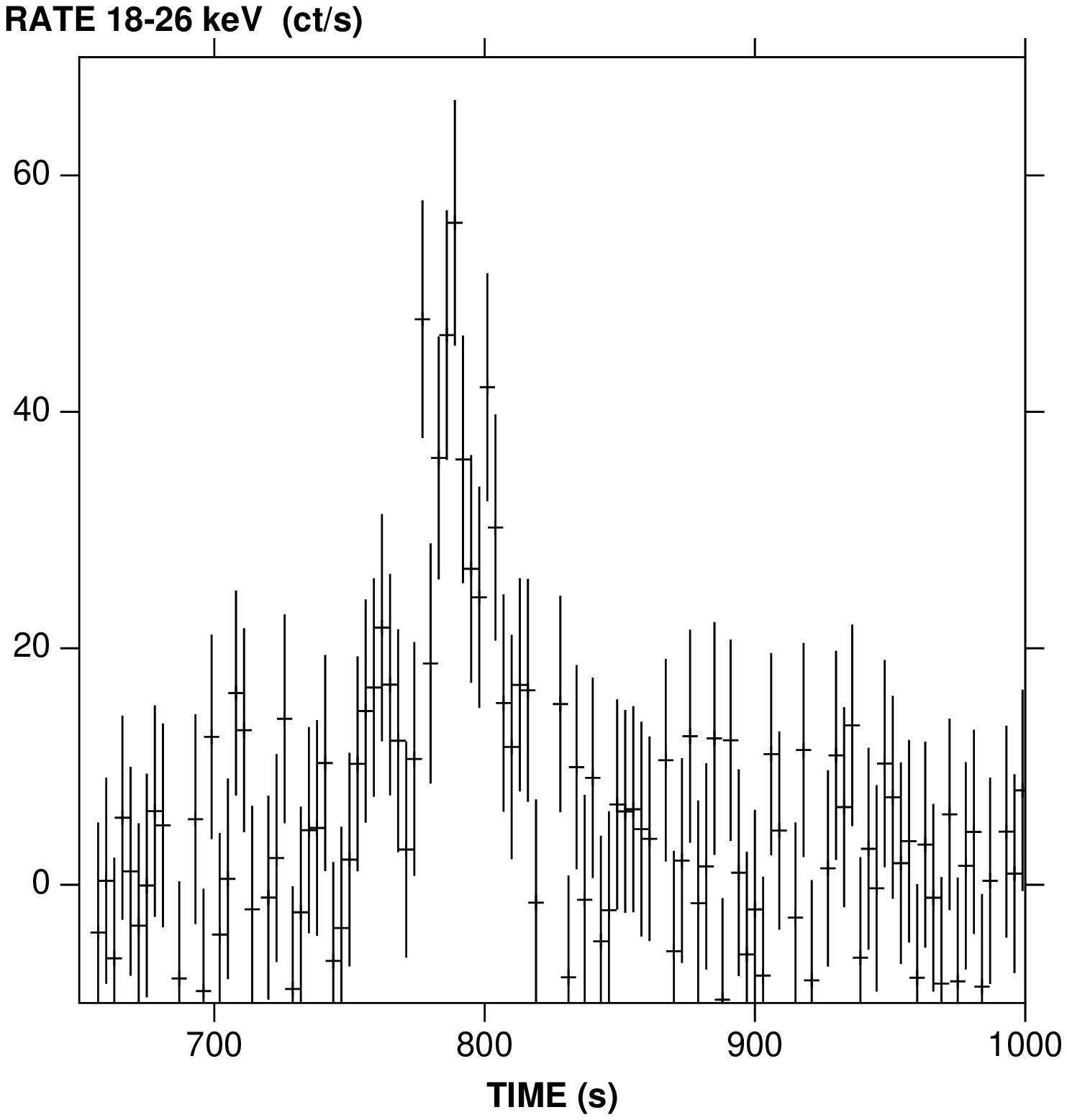}
      \caption{From the top to bottom: JEM-X1 (3-6 keV), JEM-X1 (6-10 keV) and
       ISGRI (18-26 keV) lightcurves including the type-I X-ray 
burst from \igr (binsize = 3~s). 
The \integral\ persistent flux is 
       consistent with zero.}
         \label{fig:light}
   \end{figure}
%

\section{Results}

\subsection{The X--ray burst caught with JEM-X and IBIS/ISGRI}

We show in Fig.~\ref{fig:ibis_ima} the 20-25 keV IBIS/ISGRI image collected
during the \igr\ activity, namely during the X--ray burst.
The persistent emission is below the detector level as can be inferred   
by the temporal profiles of \igr\ shown in Fig.~\ref{fig:light}.
A 2$\sigma$ IBIS upper limit (1.5 Ms exposure time) of 
3$\times$10$^{-12}$ erg~cm$^{-2}$~s$^{-1}$ to the 20-40 keV persistent flux can be derived (\cite{bird:06}).

The decay times of the burst are 71.6~s and 7.3~s in the 3--6 kev and 20-25 keV energy
bands  respectively, clearly indicative of the spectral softening typical 
of type-I X--ray bursts.
There is also evidence of a double-peaked burst profile, signature of
Eddington limited bursts showing photospheric radius expansion.

\begin{figure*}[!t]
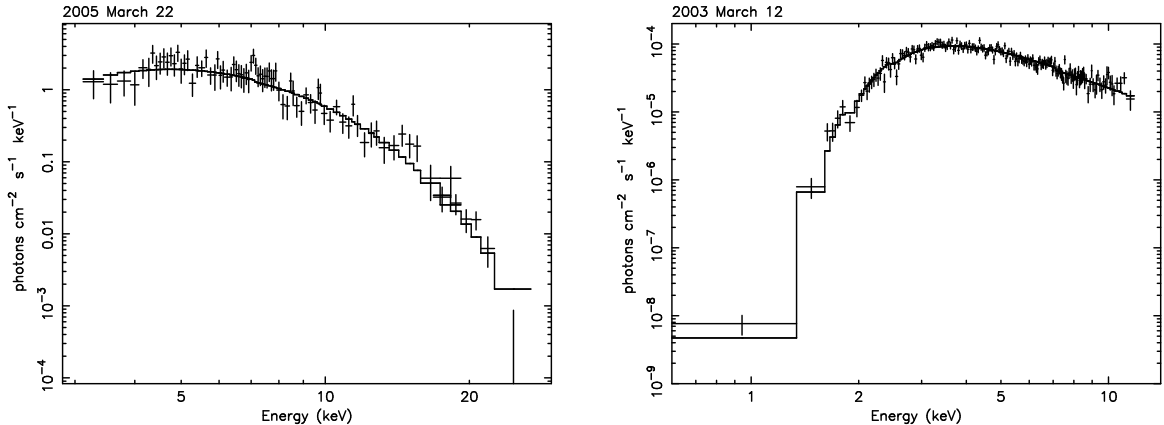

\hbox{\hspace{0.5cm}
\includegraphics[height=7.2cm,angle=-90]{delsanto07_fig5.ps}
\vspace{1.0cm}
\hspace{.6cm}
\includegraphics[height=7.2cm,angle=-90]{delsanto07_fig6.ps}}
\vspace{-1.0cm}
\caption[]{Combined IBIS/ISGRI and JEM-X1 photons spectra and model (black-body) of the burst detected by \integral\ ({\it{left}}); 
spectrum of the persistent emission of \src\ 
fitted with a simple absorbed power-law (EPIC/PN camera observation) is shown ({\it{right}}).
}
\label{fig:spec}
\end{figure*}

Combining the JEM-X and IBIS burst averaged spectra, the best-fit ($\chi^2$=80.9/79 d.o.f.) is achieved 
with an absorbed black-body model with kT = 1.8$\pm$0.1 keV and N$_{H}$=6.4$_{-2.7}^{+10.5}$~$\times10^{22}$~cm$^{-2}$
(Fig. ~\ref{fig:spec}, {\it{left}}). 
In the energy range 1-30 keV, the unabsorbed flux is  2.6$^{+1.7}_{-1.0}$ $\times$10$^{-7}$~erg~cm$^{-2}$~s$^{-1}$. 
Following Kuulkers et al. (2003), we assume an Eddington luminosity corresponding to 3.8$\times$10$^{38}$ erg~s$^{-1}$,
thus deriving a source distance in the range of 2.8--4.6~kpc.

Moreover, Kuulkers et al. (2003) report on few type-I bursts showing higher peak luminosities,
up to 7$\times$10$^{38}$ erg~s$^{-1}$.
If we assume such a high luminosity,
we then obtain a distance $\sim$6~kpc, which corresponds better to the observed high absorption.
On the other hand, absorption intrinsic to the source 
can be invoked. 
Blackbody normalization parameter translates into an emitting sphere radius 
between 10 and 25 km for a source distance of 3 kpc, and 20-50 km at 6 kpc.
We favour a lower value of the distance, 3~kpc, to be consistent with neutron star radii.

\subsection{\xmm\ and \chandra}
\begin{figure}[t!]
  \centering
   \includegraphics[angle=0,height=8.1cm]{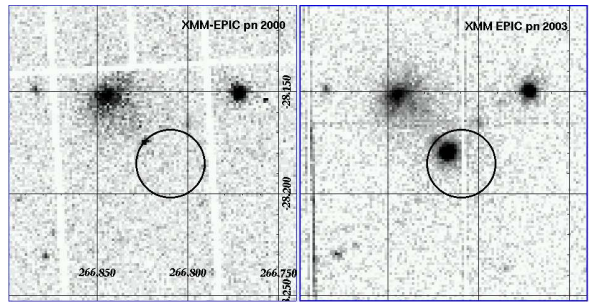}
     \vspace{-4.1cm}
     \caption{X--ray images (0.5-10 keV) of the source field obtained
      with the EPIC-PN camera in 2000 (on the left) and in 2003 (on the right). The transient \src\
is clearly evident in the centre of the 2003 image. 
The circle (1$'$ radius) marks the uncertainty region of the burst observed with JEM-X.
}
\label{fig:ima}
   \end{figure}

In order to show the transient nature of \src, a close-up view of the EPIC 0.5--10 keV
images from the 2000 and 2003 observations is shown (Fig.~\ref{fig:ima}).
The solid circle marks the JEM-X error box of the burst (\cite{brandt:06}).

\begin{figure}[t!]
  \centering
   \includegraphics[angle=0,height=5.0cm]{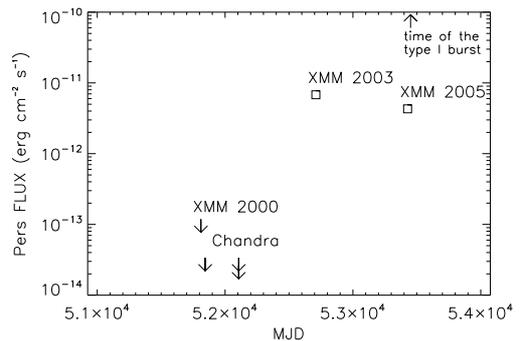}
      \caption{Long term X--ray lightcurve of \src. The two observations
performed with \xmm\ in 2003 and 2005 are marked with squares, while  meaningful upper limits
are reported with arrows. 
The persistent flux (2--10 keV) is corrected for the
absorption. Uncertainties on fluxes are smaller than symbols used.}
\label{fig:lc}
   \end{figure}

Detections and upper-limits estimated during
\chandra\ and \xmm\ observations are shown in Fig.~\ref{fig:lc}.
Our object displays a dynamic range of at least two orders of magnitude.
\chandra's upper limits to the unabsorbed fluxes (0.5-10 keV) have been 
obtained from estimated ACIS 3$\sigma$ upper limits to the source count rate,
assuming a power-law spectrum with $\Gamma$=2 and an
absorbing column density of 6 $\times10^{22}$~cm$^{-2}$.

The best \xmm\ spectrum was obtained during the 2003 observation (Fig. ~\ref{fig:spec}, {\it{right}}),
because of a favourable position (almost on-axis) in the field of view, compared to the 
$\sim$13$'$ off-axis distance during the 2005 pointing.
The fit of the 2003 spectrum with an absorbed power-law already resulted in a good fit to the data 
($\chi^2$=205.9/201 d.o.f.): a photon index of 2.1$\pm${0.1},
a column density of (8.9$\pm${0.5})~$\times10^{22}$~cm$^{-2}$, and a
2--10 keV flux corrected for the absorption
of $(6.8 \pm {0.4}) \times$10$^{-12}$~erg~cm$^{-2}$~s$^{-1}$ have been obtained.

The fit with an absorbed power-law to the 2005 spectrum resulted in similar parameters
and in an unabsorbed flux  of $4.3 ^{+0.4} _{-1.0} \times$10$^{-12}$~erg~cm$^{-2}$~s$^{-1}$  (2--10~keV).
A proper timing analysis could not be performed because of several gaps 
in the light curves caused by the high background events rejection.


\section{Discussion and Conclusions}
\label{sect:discussion}

The discovery of a type-I X--ray burst with \integral\ from a source
positionally coincident, within the errors, with \src, allow us to identify
the nature of this faint transient as a Low Mass X--ray Binary containing a neutron star,
and to determine the source distance ($\sim$3~kpc).
Furthermore, we can use the known properties of type-I bursters to derive some constraints
on the accretion history of \src, as follows. 

It is known that the quantity $\alpha$=L$_{pers}$$\times$t$_{rec}$/E$_{burst}$ 
(where L$_{pers}$ is the
persistent luminosity, t$_{rec}$ is the burst recurrence time and E$_{burst}$ the energy emitted
during the burst) is usually in the range  40--100 (see e.g. Strohmayer \& Bildsten, 2006). 
The \src\ burst is Eddington limited and displays a decay time of 70~s, implying 
E$_{burst}$$\sim$$2.7\times10^{40}$~erg.
Assuming for L$_{pers}$ the value of $10^{34}$~erg~s$^{-1}$, as observed in 2005 with \xmm,
we derive  t$_{rec}$$\sim$3$\times$($\alpha$/40)~yrs. 
This implies that the source luminosity between the two XMM-Newton observations remained at the same
level, indicating that 
the two \xmm\ observations performed in 2003 and 2005
caught the same outburst. 

However, we are aware that the range for  $\alpha$ as above  
was derived for much more luminous sources 
(see e.g. the sample in van Paradijs et al. 1988).
Indeed, the burst properties for faint sources (such as \src), those showing in outburst 
accretion rates smaller than 10$^{-12}$~M$_\odot$~yr$^{-1}$, are still unknown, and could be different. 
Moreover, we cannot be sure that $10^{34}$~erg~s$^{-1}$ was the maximum source luminosity 
reached by \src\ in the years preceding the burst.
On the other hand, if L$_{pers}$ were much less than $10^{34}$~erg~s$^{-1}$, the time needed to produce
the type-I burst would have been much longer than that allowed by the upper limits placed by
the 2000 observations.

\src\ is a member of the class of the Very Faint X--ray Transients 
(VFXTs; King \& Wijnands 2006, Wijnands et al. 2006),
where the peak outburst luminosity is in the range 10$^{34}$--10$^{36}$~erg~s$^{-1}$, 
almost three order of magnitudes fainter than outbursts from ``typical'' Galactic X--ray transients.
During the  quiescent emission the \src\ luminosity drops  
below $\sim$5$\times$10$^{32}$~erg~s$^{-1}$.
These low luminosity transients have been mainly discovered with high sensitivity instruments,
during $Chandra$ and \xmm\ surveys of the GC region (Wijnands et al. 2006, and references therein). 
\src\ is not unique among  VFXTs in displaying type-I X--ray bursts
(Hands et al. 2004, Cornelisse et al. 2002b). These may likely be the same class 
of the ``burst-only'' sources observed in the GC region 
with the Wide Field Cameras on-board BeppoSAX satellite (Cocchi et al. 2001, Cornelisse et al. 2002a).
However, the persistent flux was below the sensitivity threshold of the WFCs ($<$10$^{36}$~erg~s$^{-1}$). 
\src\ could be the first VFXT with ``quasi-persistent'' outbursts, similar
to the brighter transient LMXRBs which displays outbursts lasting few years (e.g. MXB~1659-29 and
KS~1731-260, \cite{cackett:06}).
\src\ might not be unique in this respect 
and other VFXTs could possibly display long-lived outbursts, 
but the available 
observations are probably too sparse to demonstrate it.

Previously, the $Chandra$ and \xmm\ surveys of the Galactic Centre region 
found that the VFXTs are mainly concentrated
near the GC direction, suggesting that the high stellar density near Sgr~A$^{*}$ could play a role in the
formation of these faint transients (e.g. King 2000). 
The result of our analysis that the source distance is $\sim$3~kpc
casts some doubt on the distribution of the VFXTs in general, as already suggested by Wijnands et al. 2006, from the 
location of SAX~J1828.5--1037 (another VFXT, see Hands et al. 2004).


\begin{acknowledgements}
Based on observations obtained with \xmm\ and \integral, ESA science
missions with instruments and contributions directly funded by ESA
member states and the USA (NASA). 
Data analysis is supported by the Italian Space Agency (ASI)
through contract ASI/INAF I/023/05/0.
\end{acknowledgements}


\end{document}